\documentstyle[amssymb,preprint,aps,prl,epsf]{revtex}

\begin{document}
\draft
\title{Normal ordering solution to quantum dissipation and its induced decoherence }
\author{G.R. Jin, D.L. Zhou, Yu-xi Liu$^{\dag}$, X.X. Yi and C.P. Sun\cite{email,www}}
\address{Institute of Theoretical Physics,Academia Sinica,Beijing, 100080,China\\
$^{\dag}$The Graduate University for Advanced Studies (SOKEN),\\
Hayama, Kanagawa, 240-0193, Japan\\
\medskip}
\author{\parbox{14.2cm}{\small\hspace*{3mm}We implement the normal ordering
technique to study the quantum dissipation of a single mode
harmonic oscillator system. The dynamic evolution of the system is
investigated for a reasonable initial state by solving the
Schr\"{o}dinger equation directly through the normal ordering
technique. The decoherence process of the system for the cases
$T=0K$ and $T\neq0K$ is investigated as an application. \vskip30pt
PACS numbers: 42.50 Fx,03.65-Bz}}
\maketitle

\flushbottom \narrowtext \vskip20pt

\newpage

\section{Introduction}

The quantum dissipation of a single mode harmonic oscillator coupled to a
reservoir with an extremely large number of degrees of freedom has been
intensely studied \cite{Louisell,Cal&Legg,Lax,Ullersma,Ford} in the past
decades with various methods, such as master-equation approach, Langevin
approach. By using the Markovian master-equation approach, the effect of
dissipation on a macroscopic superposition of two coherent states of a
harmonic oscillator was studied \cite{Walls&Milburn}. It shows that the
superposition of the system will be transformed into a classical mixture
state due to the effects of the environment \cite{Zurek}, which is called
decoherence process. The rate of the decoherence is proportional to the
distance between the two states, which has been demonstrated in cavity QED
experiment \cite{haroche}. The wave function structure of a single mode
boson system plus a bath of many bosons is studied for both factorization
case \cite{Yu&Sun,Sun&Yu} and partial factorization case \cite{Sun&Gao}. It
shows that when the Brownian motion effect is ignored in certain conditions
the total wave function of the total system can be written in a form of a
product of the bath and system components. Based on the result \cite{Sun&Gao}
, we have studied the decoherence process of excitons in an idealized
quantum well placed in a lossy cavity \cite{Liu&Sun}. The result shows that
the coherence of excitonic superposition is reduced in an oscillating form
due to the dissipation of the optical cavity. In references \cite
{Moussa,Oliveira1}, by solving the Heisenberg equations, the state vector of
a driven-damped oscillator system evolved from an initial coherent state has
also solved explicitly.

As far as our knowledge is concerned, the dynamic evolution of a damped
oscillator system is only investigated by implementing the
Heisenberg-Schr\"{o}dinger picture transformation (HSPT) \cite
{Sun&Gao,Moussa,Oliveira1}. Can we solve the Schr\"{o}dinger equation
directly and get the state vector of a damped harmonic oscillator system
without HSPT? To answer this question, in this paper, we firstly apply
normal ordering technique (NOT) to study the time evolution of a damped
harmonic oscillator. The explicit form of the state vector of the total
system evolved from an initially factorized coherent state is calculated.
Besides, as one of the applications, we will devote ourselves to studying
the decoherence behavior of the quantum dissipating system. For zero
temperature case, the decoherence time of a superposition of two coherent
states due to dissipation is given in a explicit form which agrees with the
previous results \cite{Walls&Milburn,Liu&Sun}. The effect of finite
temperature of the reservoir on the decoherence rate is also investigated.

This paper is organized as following: in section II, we develop the normal
ordering technique to solve the dynamic evolution of a damped harmonic
oscillator under certain initial condition. In section III, the decoherence
of a superposition state of the single mode boson system is investigated. In
subsection A, the decoherence process of the system in the case of zero
temperature is studied for an factorized initial state. We find that, the
decoherence occurs in a time scale which is sensitive to the distance of the
initial superposed states. The temperature effect on the coherence evolution
of a damped oscillator is investigated in subsection B by using NOT.

\section{Normal Ordering Technique}

The normal ordering technique \cite{Louisell} is firstly introduced to study
the dynamic evolution of a driven oscillator as well as that of two weakly
coupled oscillators without dissipation. In this section we adopt the spirit
of NOT to study the dissipation process of a single-mode boson system
coupled with a reservoir.

We consider a harmonic oscillator coupled to a reservoir which involves a
large collection of systems with many degrees of freedom, that is a damped
oscillator. The reservoir is modeled as a harmonic oscillator bath, which in
practice can be the modes of the radiation or the quantized modes of elastic
vibrations (phonons) in a solid. The Hamiltonian for the system plus bath
can be described as

\begin{equation}
H=\hbar \omega a^{\dagger }a+\hbar \sum_j\omega _jb_j^{\dagger }b_j+\hbar
\sum_j(g_ja^{\dagger }b_j+H.c.),
\end{equation}
where $a(a^{\dagger })$ denotes the annihilation (creation) operator of the
harmonic oscillator with frequency $\omega $, and $b_j(b_j^{\dagger })$ is
the annihilation (creation) operator for the mode of the bath with frequency
$\omega _j$, which obey the boson commutation relations $[b_j,b_j^{\dagger
}]=\delta _{j,j^{^{\prime }}}$. $g_j$ denotes coupling constant between the
harmonic oscillator with frequency $\omega $ and the oscillator mode of the
reservoir with frequency $\omega _j$. The model described in eq.(1) can be
exactly solved as long as the coupling constant $g_j$ and the spectrum
density of the bath are specified explicitly \cite{Sun&Gao}. Here we adopt
the spirit of the normal ordering method to study the dynamic evolution of a
single mode oscillator with dissipation.

The state vector of the whole system obeys the Schr\"{o}dinger equation
which has a solution of the form $\left| \psi (t)\right\rangle =U(t)\left|
\psi (0)\right\rangle $, where the time-evolution operator $U(t)$ satisfies $%
i\hbar \partial _tU(t)=HU(t)$, with the initial condition $U(0)=1$. We
assume that evolution operator has its normal order form $U(t)=U^{(n)}(t)$.
Since the normal form of any operator is unique, one can establish the
one-to-one corresponding relationship between the normal ordered evolution
operator $U^{(n)}(t)$ and an ordinary function $\overline{U}^{(n)}(t)$, with
$\overline{U}^{(n)}(t)=\left\langle \alpha ,\left\{ \beta _j\right\} \right|
U^{(n)}(t)\left| \left\{ \beta _j\right\} ,\alpha \right\rangle $, where $%
\left| \left\{ \beta _j\right\} \right\rangle =\left| \beta _1\right\rangle
\left| \beta _2\right\rangle \cdot \cdot \cdot \left| \beta _\infty
\right\rangle $. Such a corresponding relation defines a map $\aleph ^{-1}$ ,

\begin{equation}
\aleph ^{-1}:U^{(n)}(t)\rightarrow \overline{U}^{(n)}(t)=\left\langle \alpha
,\left\{ \beta _j\right\} \right| U^{(n)}(t)\left| \left\{ \beta _j\right\}
,\alpha \right\rangle .
\end{equation}
We may also define the inverse transformation $\aleph $ ,

\begin{equation}
\aleph :\overline{U}^{(n)}(t)\rightarrow U^{(n)}(t)=U(t).
\end{equation}
One can write the Schr\"{o}dinger equation in the normal ordering form. With
the operator $\aleph ^{-1}$, one can further take diagonal coherent state
matrix elements of both sides of the normal ordered Schr\"{o}dinger equation
and get a c-number equation of $\overline{U}^{(n)}$,
\begin{eqnarray}
i\partial _t\overline{U}^{(n)} &=&[\omega \alpha ^{*}\left( \alpha +\frac %
\partial {\partial \alpha ^{*}}\right) +\sum_j\omega _j\beta _j^{*}\left(
\beta _j+\frac \partial {\partial \beta _j^{*}}\right) +\sum_jg_j  \nonumber
\\
&&\times \alpha ^{*}\left( \beta _j+\frac \partial {\partial \beta _j^{*}}%
\right) +\sum_jg_j^{*}\beta _j^{*}\left( \alpha +\frac \partial {\partial
\alpha ^{*}}\right) ]\overline{U}^{(n)}.
\end{eqnarray}
Now we need to solve the equation for $\overline{U}^{(n)}(t)$ with the
initial condition $\overline{U}^{(n)}(0)=1$.

We let

\begin{equation}
\overline{U}^{(n)}=\exp [A\alpha ^{*}\alpha +\sum_jB_j\beta _j^{*}\beta _j+{%
\sum_{j,j^{\prime }}}^{\prime }B_{j,j^{^{\prime }}}\beta _j^{*}\beta
_{j^{^{\prime }}}+\sum_jC_j\beta _j^{*}\alpha +\sum_jD_j\alpha ^{*}\beta _j],
\end{equation}
Where the prime in ${\sum }^{\prime }$ denotes sum over index $``j"$ and $%
``j^{^{\prime }}"$ with the condition $j\neq j^{^{\prime }}$. Substituting
eq.(5) into eq.(4), we get the coupled equations for the time-dependent
coefficients

\begin{equation}
\dot{B}_j=-i\omega _j(1+B_j),  \eqnum{6.a}
\end{equation}

\begin{equation}
\dot{C}_j=-i\omega _jC_j-ig_j^{*}(1+A),  \eqnum{6.b}
\end{equation}

\begin{equation}
\dot{D}_j=-i\omega D_j-ig_j(1+B_j)-i{\sum_{j^{\prime }}}^{\prime
}g_{j^{^{\prime }}}B_{j^{^{\prime }},j},  \eqnum{6.c}
\end{equation}

\begin{equation}
\dot{A}=-i\omega (1+A)-i\sum_jg_jC_j,  \eqnum{6.d}
\end{equation}

\begin{equation}
\dot{B}_{j,j^{^{\prime }}}=-i\omega _jB_{j,j^{^{\prime
}}}-ig_j^{*}D_{j^{^{\prime }}},  \eqnum{6.e}
\end{equation}
with the initial condition $A(0)=B_{j,j^{^{\prime }}}(0)=C_j(0)=D_j(0)=0$.
The solution of eqs.(6) is easily obtained by using Wigner-Weisskopf
approximation \cite{Louisell},

\begin{eqnarray}
A(t)=u(t)-1, &&B_j=e^{-i\omega _jt}-1,  \eqnum{7.a} \\
C_j(t)=u_j(t), &&D_j(t)=v_j(t),  \eqnum{7.b} \\
B_{j,j^{^{\prime }}}(t) &=&v_{j,j^{^{\prime }}}(t),  \eqnum{7.c}
\end{eqnarray}
where, for the bath with the frequency spectrum distribution $\rho (\omega )$

\begin{equation}
u(t)=e^{-i\widetilde{\omega }t}e^{-\gamma t/2},  \eqnum{8.a}
\end{equation}

\begin{equation}
u_j(t)=-g_j^{*}e^{-i\omega _jt}\frac{e^{i(\omega _j-\widetilde{\omega }%
)t}e^{-\gamma t/2}-1}{\omega _j-\widetilde{\omega }+i\gamma /2},  \eqnum{8.b}
\end{equation}

\begin{equation}
v_j(t)=-g_je^{-i\omega _jt}\frac{e^{i(\omega _j-\widetilde{\omega }%
)t}e^{-\gamma t/2}-1}{\omega _j-\widetilde{\omega }+i\gamma /2},  \eqnum{8.c}
\end{equation}

\begin{equation}
v_{j,j^{^{\prime }}}(t)=\frac{g_j^{*}g_{j^{^{\prime }}}e^{-i\omega _jt}}{%
\omega _{j^{^{\prime }}}-\widetilde{\omega }+i\gamma /2}\left( \frac{%
e^{i(\omega _j-\widetilde{\omega })t}e^{-\gamma t/2}-1}{\omega _j-\widetilde{%
\omega }+i\gamma /2}-\frac{e^{i(\omega _j-\omega _{j^{^{\prime }}})t}-1}{%
\omega _j-\omega _{j^{^{\prime }}}}\right) ,  \eqnum{8.d}
\end{equation}
here, $\gamma =2\pi \rho (\omega )\left| g(\omega )\right| ^2$ is the
damping constant of the oscillator induced by the coupling to the
environment, and $\widetilde{\omega }=\omega +\Delta \omega $ is the
renormalized physical frequency , with $\Delta \omega $ is the lamb
frequency shift

\begin{equation}
\Delta \omega =-P\int_0^\infty d\omega _j\frac{\rho (\omega _j)\left|
g(\omega _j)\right| ^2}{\omega _j-\omega },  \eqnum{9}
\end{equation}
where ``$P$'' denotes the Cauthy principle part.

If, as an example, the initial state of the total system is

\begin{equation}
\left| \psi (0)\right\rangle =\left| \alpha \right\rangle \otimes \left|
\left\{ \beta _j\right\} \right\rangle ,  \eqnum{10}
\end{equation}
where $\left| \left\{ \beta _j\right\} \right\rangle =\prod_j\left| \beta
_j\right\rangle $ denotes multimode coherent state of the bath. Then at any
time $t$ the total system will evolve into

\begin{equation}
\left| \psi (t)\right\rangle =\left| \alpha u(t)+\sum_j\beta
_jv_j(t)\right\rangle \otimes \left| \left\{ \beta _je^{-i\omega _jt}+\alpha
u_j(t)+{\sum_{j^{^{\prime }}}}^{\prime }v_{j,j^{^{\prime }}}(t)\beta
_{j^{^{\prime }}}\right\} \right\rangle ,  \eqnum{11}
\end{equation}
where we have used the sum rule,
\begin{equation}
|\alpha u(t)+\sum_j\beta _jv_j(t)|^2+\sum_j|\beta _je^{-i\omega _jt}+\alpha
u_j(t)+{\sum_{j^{^{\prime }}}}^{\prime }v_{j,j^{^{\prime }}}(t)\beta
_{j^{^{\prime }}}|^2=|\alpha |^2+\sum_j|\beta _j|^2,  \eqnum{12}
\end{equation}
which is derived from the normalized condition $\langle \psi (t)|\psi
(t)\rangle =1$. We see from eq.(13), due to the bath fluctuation and the
back-action of system on the bath, the state vector evolved from factorized
initial state becomes fully entangled. If the Brownian effect caused by the
terms $\sum_j\beta _jv_j(t)$ can be ignored, the total state vector can be
partially factorized \cite{Sun&Gao}. We can further consider $T=0K$ for the
bath, that is, all the oscillator modes of the reservoir are in vacuum state
initially. In cavity QED, this means that we neglect the background
radiation of cavity. Then from eq.(13), the state vector can be simplified as

\begin{equation}
\left| \psi (t)\right\rangle =\left| \alpha u(t)\right\rangle \otimes \left|
\left\{ \alpha u_j(t)\right\} \right\rangle ,  \eqnum{13}
\end{equation}
and the corresponding sum rule is

\begin{equation}
|u(t)|^2+\sum_j|u_j(t)|^2=1.  \eqnum{14}
\end{equation}

\section{ Decoherence of superposition state of the harmonic oscillator due
to dissipation}

Recently the mesoscopic superposition state of the cavity mode states \cite
{haroche} has been prepared experimentally. The phase shift between two
components of the superposition can be well controlled by adjusting the
interaction time between atoms and the cavity mode. Then the even and odd
Schr\"{o}dinger cats of cavity states were prepared in the experiment, with
which a new scheme for logic qubit encoding are proposed \cite{Oliveira2} to
simplify the error correction circuits and improve the efficiency of the
error correction in quantum computation. Decoherence process of the
superposed cavity state was also demonstrated in the experiment. Due to the
presence of dissipation in the cavity, the coherent information of
Schr\"{o}dinger cat state will be lost within a time scale which is
sensitive to the size of the meter states.

As an application, in this section we will study the decoherence behavior of
the damped oscillator described by eq.(1). Starting with eq.(11), we will
calculate the decoherence factor, which is defined as the coefficients of
the off-diagonal element of the reduced density operator of system, for both
$T=0K$ and finite temperature of the heat bath.

\subsection{Zero-temperature case}

A single mode boson system coupled to a bath which is composed of a
collection of many harmonic oscillators will result in the decohence process
of the system, that is, a superposition state of the system will be
transformed into a statistical mixture state due to the influence of the
bath. If the initial state of the total system is
\begin{equation}
|\Psi (0)\rangle =(C_1|\alpha _1\rangle +C_2|\alpha _2\rangle )\otimes
\left| \left\{ 0_j\right\} \right\rangle ,  \eqnum{15}
\end{equation}
that is, a superposition state for the system, and all the oscillator modes
of the bath are in the vacuum states $\left| \left\{ 0_j\right\}
\right\rangle $ initially. Therefore from eq.(15), the state vector of the
total system at time $t$ can be written as
\begin{equation}
|\Psi (t)\rangle =C_1|\alpha _1u(t)\rangle \otimes |\left\{ \alpha
_1u_j(t)\right\} \rangle +C_2|\alpha _2u(t)\rangle \otimes |\left\{ \alpha
_2u_j(t)\right\} \rangle ,  \eqnum{16}
\end{equation}
where the explicit form of $u(t)$, $u_j(t)$ are given in eq.(8), and
eq.(11.a), respectively. In addition, the sum rule

\begin{equation}
\sum_j|u_j(t)|^2=1-|u(t)|^2,  \eqnum{17}
\end{equation}
has been used in eq.(16), which is derived from the condition,
\begin{equation}
\langle \Psi (t)|\Psi (t)\rangle =|C_1|^2+|C_2|^2+C_1C_2^{*}\langle \alpha
_2|\alpha _1\rangle +C_1^{*}C_2\langle \alpha _1|\alpha _2\rangle .
\eqnum{18}
\end{equation}

The dynamic decoherence process can be investigated quantitatively by
calculating the reduced density matrix $Tr_R(|\Psi (t)\rangle \langle \Psi
(t)|)$ of the system at any time $t$. Then we will get the decoherence
factor, which is defined as the coefficient of the non-diagonal elements of
the reduced density matrix, such as
\begin{eqnarray}
F(t) &=&\prod_j\langle \alpha _1u_j(t)|\alpha _2u_j(t)\rangle  \nonumber \\
&=&e^{(-\frac 12|\alpha _1|^2-\frac 12|\alpha _2|^2+\alpha _1^{*}\alpha
_2)\sum_j|u_j(t)|^2}.  \eqnum{19}
\end{eqnarray}
By using the sum rule of eq.(17), the decoherence factor becomes
\begin{equation}
F(t)=e^{(-\frac 12|\alpha _1|^2-\frac 12|\alpha _2|^2+\alpha _1^{*}\alpha
_2)(1-|u(t)|^2)}.  \eqnum{20}
\end{equation}
The characteristic time $\tau _d$ of the decoherence of the superposition
state is determined by the short time behavior of $|F(t)|$, that is, the
norm of the decoherence factor. Within the time scale $\gamma t\ll 1$, the
decoherence factor can be simplified as following,

\begin{equation}
F(t)=e^{(-\frac 12|\alpha _1|^2-\frac 12|\alpha _2|^2+\alpha _1^{*}\alpha
_2)\gamma t}.  \eqnum{21}
\end{equation}
If we consider that $\alpha _1=\alpha $, and $\alpha _2=\alpha e^{i\Delta
\varphi }$, where $\Delta \varphi $ is the phase shift of the initial
superposed states. Then the characteristic time is determined as following,

\begin{equation}
\tau _d^{-1}=2|\alpha |^2\gamma \sin ^2\left( \Delta \varphi /2\right) ,
\eqnum{22}
\end{equation}
where $|\alpha |^2$ is the mean number of the oscillator. We can define the
``distance'' $D=|\alpha _1-\alpha _2|=2|\alpha |\sin \left( \Delta \varphi
/2\right) $ between the two superposed states. Substituting $D$ into
eq.(22), we obtain that the characteristic time $\tau _d=\frac{2\tau _p}{D^2}
$, where $\tau _p=1/\gamma $ is the life time of the oscillator due to its
energy dissipation. Our result shows that the decoherence time is determined
by both the phase difference of the meter states of initial superposition
state and the mean number of the quanta of the single mode boson field for a
fixed damping rate. For a special case $\Delta \varphi =\pi $, which means
that the system is prepared initially in an odd and even coherent states,
the norm of decoherence factor of eq.(20) becomes

\begin{equation}
|F(t)|=e^{-2|\alpha |^2(1-e^{-\gamma t})},  \eqnum{23}
\end{equation}
as a function of time. It shows that the coherence of the oscillator will
decrease in the exponential decay rule.

\subsection{ finite temperature case}

We assume that every oscillator mode of the bath is initially in thermal
equilibrium state, and the single-mode boson is in a superposition of two
coherent states. Then the density operator of total system at $t=0$ reads,

\begin{equation}
\rho (0)=\rho _s\otimes \rho _b,  \eqnum{24}
\end{equation}
with

\begin{equation}
\rho _s=(C_1|\alpha _1\rangle +C_2|\alpha _2\rangle )(\left\langle \alpha
_2\right| C_2^{*}+\left\langle \alpha _1\right| C_1^{*}),  \eqnum{25a}
\end{equation}
and

\begin{equation}
\rho _b=\int \prod_jd^2\beta _j\frac 1{\pi \langle n_j\rangle }e^{-|\beta
_j|^2/\langle n_j\rangle }\left| \beta _j\right\rangle \left\langle \beta
_j\right| ,  \eqnum{25.b}
\end{equation}
where $\langle n_j\rangle $ denotes the mean occupation of the $j$-th
oscillator mode with frequency $\omega _j$ of the heat bath. From eq.(11),
the decoherence factor becomes

\begin{equation}
F(t)=e^{(-\frac 12|\alpha _1|^2-\frac 12|\alpha _2|^2+\alpha _1^{*}\alpha
_2)(1-|u(t)|^2)}\prod_jf_j,  \eqnum{26}
\end{equation}
with

\begin{equation}
f_j=\int \frac{d^2\beta _j}{\pi \langle n_j\rangle }e^{-|\beta _j|^2/\langle
n_j\rangle }e^{(\alpha _2^{*}-\alpha _1^{*})u^{*}(t)v_j(t)\beta _j-C.c.}.
\eqnum{27}
\end{equation}
By using the identity

\begin{equation}
\frac 1\pi \int d^2\beta \exp (-\lambda |\beta |^2+\mu \beta +\nu \beta
^{*})=\frac 1\lambda \exp \left( \frac{\mu \nu }\lambda \right) ,  \eqnum{28}
\end{equation}
with condition $%
\mathop{\rm Re}%
\left\{ \lambda \right\} >0$ and arbitrary $\mu $ and $\nu $, we obtain $f_j$

\begin{equation}
f_j=e^{-\frac 14|\alpha _2-\alpha _1|^2|u(t)|^2|v_j(t)|^2\langle n_j\rangle
}.  \eqnum{29}
\end{equation}
Substituting $f_j$ of eq.(29) into eq.(26), we get

\begin{eqnarray}
F(t) &=&e^{(-\frac 12|\alpha _1|^2-\frac 12|\alpha _2|^2+\alpha _1^{*}\alpha
_2)(1-|u(t)|^2)}e^{-\frac 14|\alpha _2-\alpha
_1|^2|u(t)|^2\sum_j|v_j(t)|^2\langle n_j\rangle }  \nonumber \\
&=&e^{(-\frac 12|\alpha _1|^2-\frac 12|\alpha _2|^2+\alpha _1^{*}\alpha
_2)(1-|u(t)|^2)}e^{-\frac 14|\alpha _2-\alpha _1|^2|u(t)|^2\overline{n}%
(1-e^{-\gamma t})},  \eqnum{30}
\end{eqnarray}
where we have used the relation \cite{Moussa},

\begin{equation}
\sum_j|v_j(t)|^2\langle n_j\rangle =\overline{n}(1-e^{-\gamma t}),
\eqnum{31}
\end{equation}
here $\overline{n}=\left( e^{\frac{\hbar \omega }{k_BT}}-1\right) ^{-1}$,
with $k_B$ is the Boltzmann constant and $T$ is the absolute temperature of
heat bath. For low temperature, $\overline{n}\thicksim 0$, the decoherence
factor of eq.(30) will go back eq.(20). While at high temperature, $%
\overline{n}\thicksim \frac{k_BT}{\hbar \omega }$, as the previous
subsection, the characteristic time $\tau _d$ of the decoherence of the
superposition state is determined by calculating the norm of the decoherence
factor within the time scale $\gamma t\ll 1$, that is,

\begin{equation}
F(t)=e^{(-\frac 12|\alpha _1|^2-\frac 12|\alpha _2|^2+\alpha _1^{*}\alpha
_2)\gamma t}e^{-\frac 14\frac{k_BT}{\hbar \omega }|\alpha _2-\alpha
_1|^2\gamma t}.  \eqnum{32}
\end{equation}
For $\alpha _1=\alpha $, and $\alpha _2=\alpha e^{i\Delta \varphi }$, with $%
\Delta \varphi $ is the phase shift between the ``meter'' states of the
initial superposition state. Then the decoherence time is determined as

\begin{equation}
\tau _d^{-1}=2|\alpha |^2\gamma \left( 1+\frac{k_BT}{2\hbar \omega }\right)
\sin ^2\left( \Delta \varphi /2\right) .  \eqnum{33}
\end{equation}
As the previous subsection we substitute the ``distance'' $D=|\alpha
_1-\alpha _2|=2|\alpha |\sin \left( \Delta \varphi /2\right) $ between the
two superposed states into eq.(33) and obtain the decoherence time $\tau _d=%
\frac{2\tau _p}{D^2}\frac 1{1+\frac{k_BT}{\hbar \omega }}$. Comparing with
the zero temperature case, we find that due to the finite temperature effect
of bath the decoherence rate of the system becomes faster.

\section{Conclusion}

The quantum dissipation process of a single-mode boson immersed in a bath of
bosons is studied by using the normal ordering technique. The dynamic
evolution of the total system is obtained and is used to study the
decoherence behavior of the system both for the zero temperature case and
for the finite temperature case. Due to the influence of the dissipation the
coherence information of the system will be lost in a time scale which is
dependent on the distance of the initial superposition state. For $T\neq 0K$%
, the finite temperature effect of the bath will decrease the decoherence
time. One can note that the normal ordering method can be also used to study
the dynamic evolution of a driven-damped oscillator system.

\acknowledgements This work is supported in part by the National Foundation
of Natural Science of China. One of authors (Yu-xi Liu) is supported
partially by Japan Society for the Promotion of Science (JSPS). He is also
indebted to Professor N. Imoto for continuous encouragement in his work.

\end{document}